\documentclass[]{spie}  

\usepackage{amsmath,amsfonts,amssymb}
\usepackage{graphicx}
\usepackage{tikz}
\usepackage{bm} 

\usetikzlibrary{shapes.geometric, arrows, positioning, calc, shadows, fit, backgrounds}
\usepackage[colorlinks=true, allcolors=blue]{hyperref}
\pgfdeclarelayer{background}
\pgfsetlayers{background,main}
\title{Instrumental artifacts in photonic lantern spectroastrometry and their mitigation with PLred
}

\author[a]{Yoo Jung Kim}
\author[a]{Michael P. Fitzgerald}
\author[a]{Malena Bloom}
\author[b]{Aidan Walk}
\author[c,e]{S\'ebastien Vievard}
\author[d,e]{Miles Lucas}
\author[d,e]{Olivier Guyon}
\author[f]{Sylvestre Lacour}
\author[f]{Elsa Huby}
\author[f]{Jehanne Sarrazin}
\author[f]{Manon Lallement}
\author[e]{Julien Lozi}
\author[g]{Nemanja Jovanovic}
\affil[a]{Department of Physics \& Astronomy, 430 Portola Plaza, University of California, Los Angeles, CA 90095, USA}
\affil[b]{University of Hawai'i at Hilo, 200 W. Kāwili St., Hilo, HI, 96720, USA}
\affil[c]{Space Science and Engineering Initiative, College of Engineering, University of Hawai'i, 640 North Aohoku Place, Hilo, HI 96720, USA}
\affil[d]{Steward Observatory, University of Arizona, 933 N Cherry Ave,Tucson, AZ 85719, USA}
\affil[e]{Subaru Telescope, National Astronomical Observatory of Japan, 650 North Aohoku Place, Hilo, 96720, HI, USA}
\affil[f]{LIRA, Observatoire de Paris, Universit\'e PSL, Sorbonne Universit\'e, Universit\'e Paris Cit\'e, CY Cergy Paris Universit\'e, CNRS, 5 place Jules Janssen, Meudon, 92195, France}
\affil[g]{Department of Astronomy, California Institute of Technology, 1200 East California Boulevard, Pasadena, CA 91125, USA}

\authorinfo{Further author information: (Send correspondence to Yoo Jung Kim)\\Yoo Jung Kim: E-mail: yjkim@astro.ucla.edu}

\begin{document} 
\maketitle

\providecommand\mnras{Monthly Notices of the Royal Astronomical Society}
\providecommand\apj{The Astrophysical Journal}
\providecommand\apjl{The Astrophysical Journal, Letters}
\providecommand\apss{Astrophysics and Space Science}
\providecommand\aap{Astronomy \& Astrophysics}
\providecommand\pasp{Publications of the Astronomical Society of the Pacific}
\providecommand\nat{Nature}

\begin{abstract}
Spectroastrometry is a powerful spectral-differential technique for probing angular scales below the resolution limit, but it is also well known to be susceptible to instrumental artifacts that can mimic or obscure real signals. In photonic lantern spectroastrometry, recently demonstrated on-sky with Subaru/FIRST-PL, the astrometric signal is encoded in relative flux variations between lantern outputs rather than in centroid shifts along a slit. This changes the artifact landscape: some slit-based spectroastrometric artifacts are avoided, but new artifact mechanisms emerge, including detector nonlinearity and spectral extraction errors, which can produce spurious features on emission or absorption lines. These lessons directly informed the design of \texttt{PLred}, an open-source Python package for photonic lantern data reduction and instrument-agnostic spectral-differential image reconstruction. We describe the origin of these artifacts and the key pipeline design choices used to mitigate them.
\end{abstract}


\section{Introduction}\label{sec:intro}

Spectroastrometry is a spectral-differential technique that probes angular scales below the resolution limit by measuring wavelength-dependent changes in the center of light of an unresolved source \cite{bai98, whelan08}. If emission at different wavelengths originates from spatially distinct regions, for example because of kinematic structure in an extended rotating source, the photocenter varies across the spectral line and encodes sub-resolution spatial information.



In principle, photon-noise-limited centroid estimation can achieve extremely high precision with increasing integration time, as the precision scales with the PSF size divided by the square root of the total number of detected photons. In practice, however, spectroastrometric measurements often do not reach this limit because they are dominated by instrumental systematics that mimic or obscure the true signal. Nevertheless, spectroastrometry has become a powerful method for investigating sub-milliarcsecond scales of compact astrophysical structures such as the innermost regions of protoplanetary disks and broad-line regions (e.g., \cite{bai98, whelan08, whe04, oudmaijer08, wheelwright12, pon08, gra18, gra19}).

\begin{figure}
\centering
\includegraphics[width=1\textwidth]{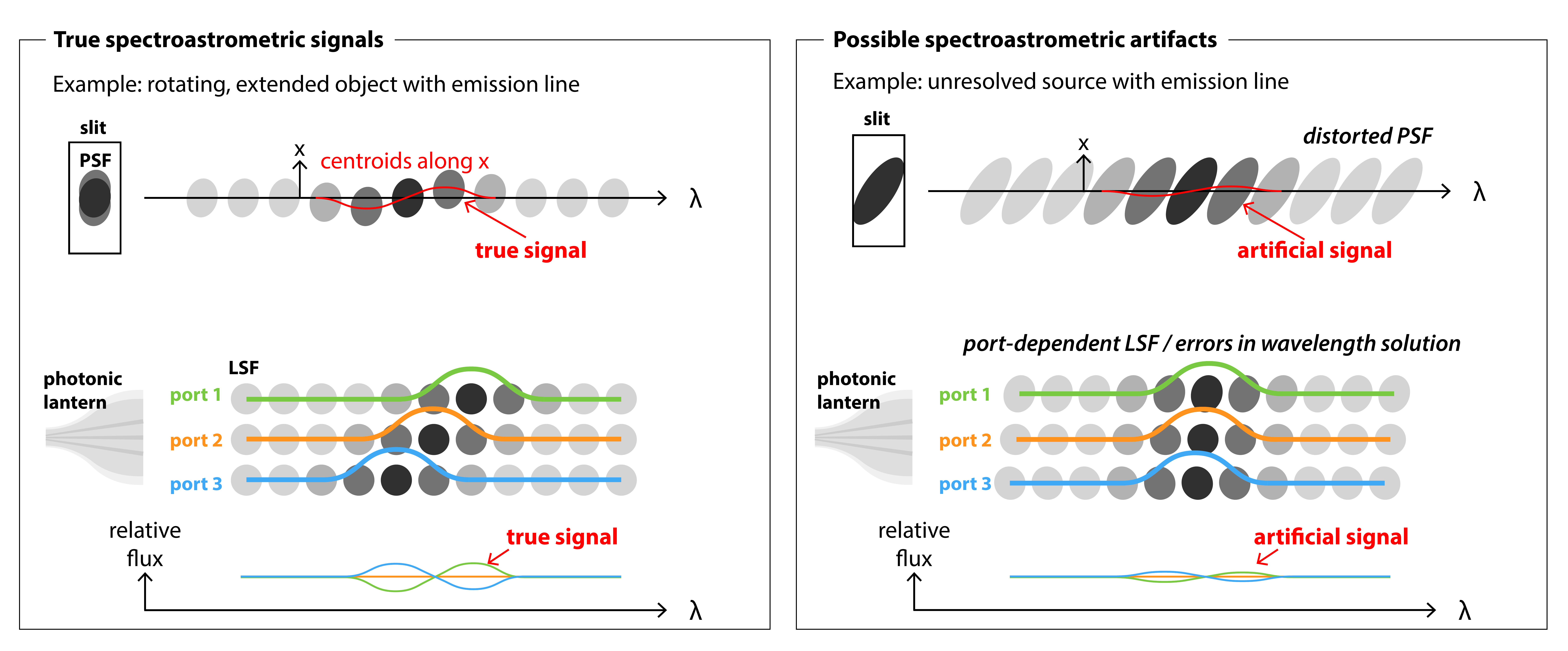}
\caption{
Illustration of true spectroastrometric signals and possible artifacts 
for traditional long-slit spectroastrometry (top rows) and photonic lantern 
spectroastrometry (bottom rows). \textbf{Left:} A rotating extended object with 
an emission line produces a genuine spectroastrometric signal: in long-slit 
spectroastrometry, the PSF centroid shifts as a function of wavelength (top); 
in photonic lantern spectroastrometry, the relative fluxes between lantern outputs 
vary across the emission line (bottom), as shown in the relative flux curves. 
\textbf{Right:} An unresolved point source with an emission line produces no true 
spectroastrometric signal, but artifacts can arise from instrumental effects. In 
long-slit spectroastrometry, a distorted PSF produces a 
spurious centroid shift on the emission line (top). In photonic lantern spectroastrometry, 
port-dependent line spread functions (LSFs) or errors in the wavelength solution 
cause the emission line to be extracted at slightly different wavelengths or with 
different profiles for each output, producing spurious relative flux variations 
that mimic a true signal (bottom).}\label{fig7:artifacts}
\end{figure}

In traditional long-slit spectroastrometry, a narrow slit isolates a portion of the point spread function (PSF),
and the centroid of the spatial profile along the slit axis is measured as a function of wavelength (e.g., top left of Fig. \ref{fig7:artifacts}).
In this configuration, the PSF falling on the slit is directly mapped onto detector pixels,
since each wavelength channel forms an image of the slit. As a result, artificial spectroastrometric signals primarily arise from uneven illumination of the slit and PSF distortions \cite{bra06, whe15}. Top right of Fig. \ref{fig7:artifacts} shows an example of such an artificial signal caused by a distorted PSF. These effects can be mitigated by using a narrower slit and repeating the observations with the slit rotated by 180$^\circ$, allowing artificial signals to be subtracted. More fundamental errors arise from detector effects such as intra-pixel sensitivity variations, which introduce a systematic noise floor in astrometric measurements in general \cite{whe08}.

A photonic lantern \cite{leon-saval05, leo10, bir15} provides a different way of encoding spatial information. It is a fiber-based mode converter that receives multimode light at the focal plane and transforms it into multiple single-mode outputs, with the spatial information distributed across the relative amplitudes of those outputs \cite{kim24_imaging}. When these outputs are fed to a spectrometer, wavelength-dependent spatial information can be measured through relative spectral variations between ports rather than through centroid shifts along a slit. See Ref. \cite{kim24_sa} for details on the concept of photonic lantern spectroastrometry.

Thus, in photonic lantern spectroastrometry, the signal is encoded in a different measurement space (see the bottom left of Fig. \ref{fig7:artifacts}). The shapes of the spectral traces do not depend on the PSF itself, but instead correspond to the images of the single-mode outputs of the photonic lantern, or the line spread functions (LSFs). As a result, the measurement is less directly sensitive to PSF-shape distortions. However, greater care is required when comparing spectra between different outputs, since relative flux measurements depend on accurate spectral extraction (bottom right of Fig. \ref{fig7:artifacts}). This effect is discussed in detail in \S\ref{sec:extraction}.

Also, detector effects manifest in a different way. In traditional slit-based astrometry, intra-pixel sensitivity variations can introduce systematic errors. In photonic lantern spectroastrometry, because the LSFs are fixed, this effect does not enter in the same way. However, detector nonlinearity can be a significant source of systematic errors, as discussed in \S\ref{sec:detector}.  
These effects were encountered in the on-sky demonstration of photonic lantern spectroastrometry with Subaru/FIRST-PL, which achieved 50 $\mu$as centroid precision in H$\alpha$ on the classical Be star $\beta$ CMi \cite{kim25}. Understanding and correcting these artifacts was essential to achieving this result, and directly motivated the design of the data reduction and image reconstruction pipeline \texttt{PLred}\footnote{\url{https://github.com/YooJung-Kim/PLred}}. 

\section{The response-map method and its implementation in PLred}\label{sec:responsemap}

\subsection{The response-map method}

The response-map method is the core framework implemented in \texttt{PLred} for recovering spectroastrometric signals from photonic lantern observations. It was developed to address the dominant practical challenge of ground-based observations: atmospheric tip-tilt jitter, which causes the PSF to wander at the lantern input and couples into the measured spectra, diluting subtle spatial signals. Details can be found in Ref. \cite{kim25}.

The response-map method uses simultaneously recorded focal-plane images to sort individual PL frames by their corresponding PSF peak position. Frames falling within the same two-dimensional PSF position bin are averaged together, building a ``response map'' for each port and wavelength channel: a two-dimensional map of how each output responds as a function of PSF position at the lantern input. 
The measured response maps $M(\lambda, i)$ 
can be modeled as a convolution of the true astronomical image with the response 
of a point source (the ``reference response map'' $M_0$):
\begin{equation}\label{eq:deconv}
M(\lambda, i) = \mathrm{Image}(\lambda) \ast M_0(\lambda, i),
\end{equation}
where $i$ is the port index. Recovering the astronomical scene thus becomes a 
deconvolution problem. Photocenter shifts appear as translations of $M$ relative 
to $M_0$, while spatially extended emission appears as broadening.

A key property of this approach is that the spectroastrometric observable, the response map, is constructed from flux variations \textit{within individual pixels} across the PSF-position grid, rather than from flux comparisons \textit{between} pixels. This distinction has important consequences for artifact mitigation. In particular, imperfections in the output imaging system that cause port-dependent LSF variations --- the artifact illustrated in the bottom right panel of Fig.~\ref{fig7:artifacts} --- do not translate as directly into spurious signals, because each pixel’s response is tracked independently across PSF positions. For the same reason, static pixel-to-pixel gain variations do not propagate directly into the astrometric measurement.


\subsection{PLred implementation}

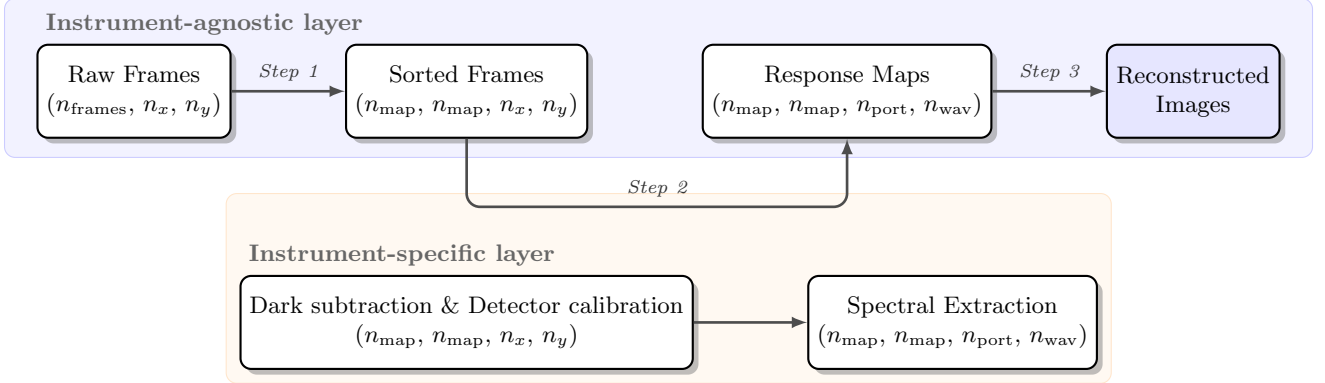
\begin{figure}[h!]
\begin{tikzpicture}[
    node distance=1.8cm and 1.5cm,
    >=latex,
    font=\small
]

\tikzstyle{block} = [
    rectangle, draw, rounded corners, 
    align=center, minimum height=3.5em, minimum width=6.5em,
    fill=white, drop shadow, line width=0.8pt
]
\tikzstyle{arrow} = [->, thick, draw=black!70, line width=1pt, rounded corners=5pt]
\tikzstyle{steplabel} = [font=\scriptsize\itshape, text=black!80, midway, above]
\tikzstyle{textnote} = [font=\footnotesize, text=black!60]

\node[block] (raw) {Raw Frames\\ ($n_{\text{frames}}$, $n_x$, $n_y$)};

\node[block, right=of raw] (mappedframes) {Sorted Frames\\ ($n_{\text{map}}$, $n_{\text{map}}$, $n_x$, $n_y$)};

\node[block, right=of mappedframes] (mappedspectra) {Response Maps\\($n_{\text{map}}$, $n_{\text{map}}$, $n_{\text{port}}$, $n_{\text{wav}}$)};

\node[block, right=of mappedspectra, fill=blue!10] (recon) {Reconstructed\\Images};

\node[block, below=1.8cm of mappedframes] (dark) {Dark subtraction \& Detector calibration\\($n_{\text{map}}$, $n_{\text{map}}$, $n_x$, $n_y$)};
\node[block, right=of dark] (extract) {Spectral Extraction\\($n_{\text{map}}$, $n_{\text{map}}$, $n_{\text{port}}$, $n_{\text{wav}}$)};

\node[textnote, anchor=south west] at (raw.north west) {\textbf{Instrument-agnostic layer}};
\node[textnote, anchor=south west] at (dark.north west) {\textbf{Instrument-specific layer}};


\draw[arrow] (raw) -- node[steplabel] {Step 1} (mappedframes);

\coordinate (turn_left) at ($(mappedframes.south)+(0,-0.9cm)$);
\coordinate (turn_right) at ($(mappedspectra.south)+(0,-0.9cm)$);

\draw[arrow] (mappedframes.south) -- (turn_left) 
    -- node[steplabel] {Step 2} (turn_right) 
    -- (mappedspectra.south);

\draw[arrow] (dark) -- node[steplabel] {} (extract);
\draw[arrow] (mappedspectra) -- node[steplabel] {Step 3} (recon);

\begin{pgfonlayer}{background}
    \node[fit=(raw)(recon)(mappedframes)(mappedspectra), 
          draw=blue!20, fill=blue!5, rounded corners, inner sep=1.2em, yshift=0.5em] (agnostic_box) {};

    \node[fit=(dark)(extract)(turn_left)(turn_right), 
          draw=orange!20, fill=orange!5, rounded corners, inner sep=0.5em] (specific_box) {};
\end{pgfonlayer}

\end{tikzpicture}
\caption{Block diagram of the \texttt{PLred} pipeline. Raw frames (shape $n_\mathrm{frames} \times n_x \times n_y$, where $n_x, n_y$ are the detector dimensions) are first sorted by PSF position into a 2D grid of $n_\mathrm{map} \times n_\mathrm{map}$ bins (Step 1). The instrument-specific layer (Step 2) applies dark subtraction and detector calibration, then extracts spectra, producing calibrated response maps of shape $n_\mathrm{map} \times n_\mathrm{map} \times n_\mathrm{port} \times n_\mathrm{wav}$, where $n_\mathrm{port}$ is the number of photonic lantern outputs and $n_\mathrm{wav}$ is the number of wavelength channels. The instrument-agnostic layer then reconstructs the astronomical image as discussed in Appendix E of Ref. \cite{kim25} (Step 3).\label{fig:plred_diagram}}
\end{figure}

\texttt{PLred} implements the response-map method presented in Ref. \cite{kim25} as a modular, two-layer pipeline. Fig. \ref{fig:plred_diagram} shows a simplified block diagram. An \textbf{instrument-agnostic layer} handles frame sorting by PSF positions (step 1) and image reconstruction (step 3; solving the deconvolution problem in Equation \ref{eq:deconv}), and can in principle be applied to any photonic lantern-fed spectrometer. An \textbf{instrument-specific layer} encapsulates the detector calibration and spectral extraction steps required to convert raw detector images into calibrated spectra suitable for response map construction (step 2). 

While the response-map method mitigates one major source of artifact, careful attention to the instrument-specific layer is required. The following section describes the two key considerations for Subaru/FIRST-PL: accurate spectral extraction (\S\ref{sec:extraction}) and detector nonlinearity correction (\S\ref{sec:detector}).

In principle, the order of steps 1 and 2 can be reversed, so that spectra are first extracted from the raw data and then sorted by PSF position. For FIRST-PL, however, the ordering adopted here was motivated by the detector nonlinearity correction, which is most robust when applied to averaged, high signal-to-noise frames.

\section{FIRST-PL-specific spectral extraction}

The FIRST-PL instrument \cite{vievard24, firstpl_poster, kim25} is a 19-port photonic-lantern fed spectrograph installed on the SCExAO instrument \cite{jovanovic15} on the Subaru Telescope. The 19 single-mode outputs are arranged in a V-groove array, forming a pseudo-slit, passes through 
a Wollaston prism to split orthogonal polarizations, and then spectrally dispersed by a volume phase holographic grating. The resulting 38 spectral traces are imaged to the Hamamatsu ORCA-Quest qCMOS detector. 
The spectrograph operates in the wavelength range of $\lambda \sim 0.6-0.8 \mu$m, with the spectral resolution of $R \sim 3,000$. In this section, we describe FIRST-PL-specific spectral extraction and calibration strategies implemented as a submodule of \texttt{PLred}.

\subsection{Necessity of accurate spectral extraction --- a forward-modeling approach}\label{sec:extraction}

In the response-map method, response maps are constructed separately for each port and wavelength, and the image reconstruction in each wavelength channel is performed by jointly finding an image that explains the response maps of all ports. Accurate wavelength solutions are therefore critical: for measuring spectroastrometric signals on narrow lines, even sub-pixel differences between ports must be taken into account. Likewise, if the LSFs are significantly distorted, spectral features can bleed differently into neighboring pixels, particularly in the presence of prominent emission or absorption lines.

\begin{figure} 
\centering
\includegraphics[width=1\textwidth]{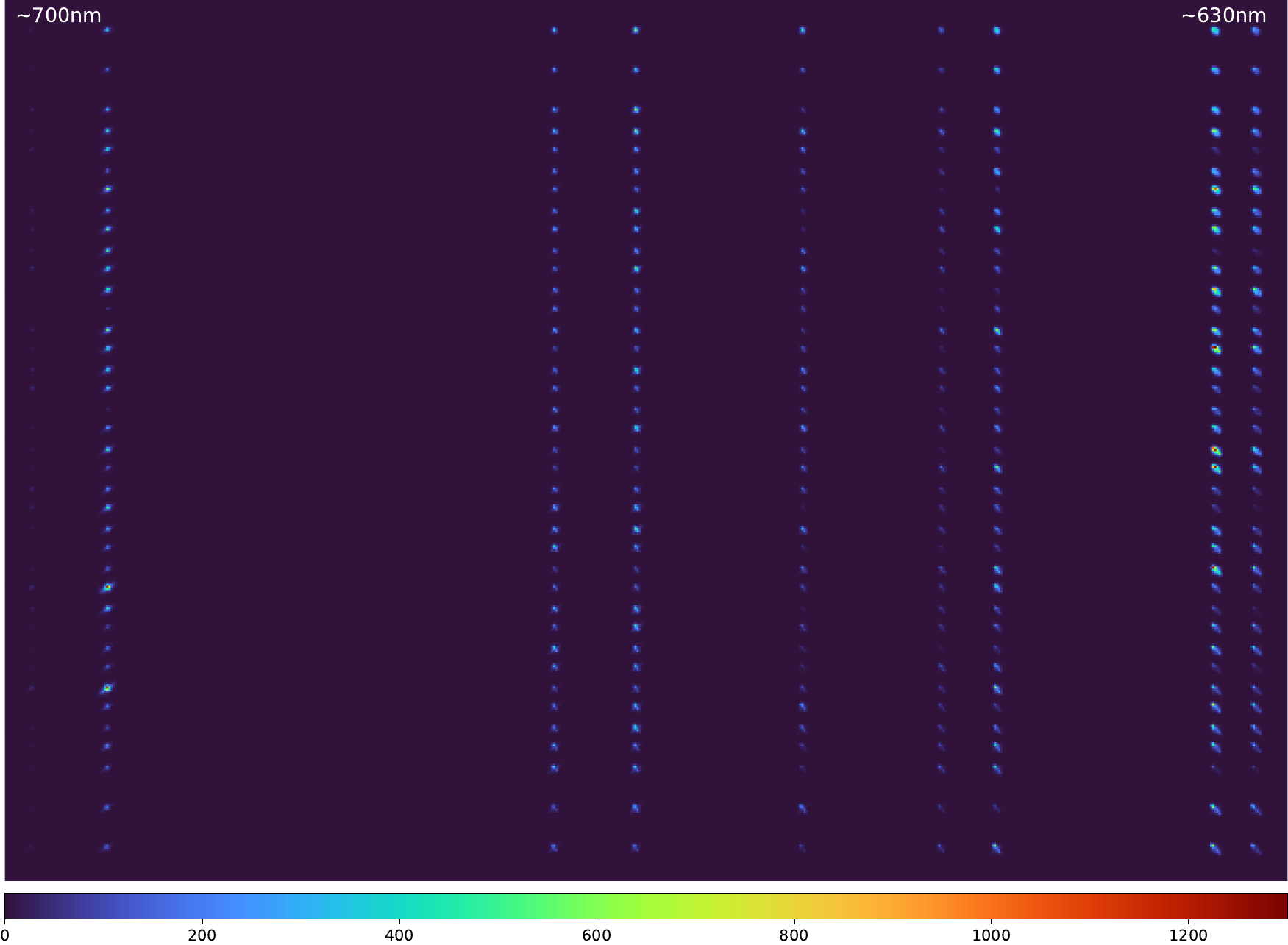}
\caption{A portion of FIRST-PL Neon lamp calibration image. The LSF morphology varies with both wavelength and port. These empirical LSFs are used in the forward modeling of spectral traces on the detector.}\label{fig7:neon}
\end{figure}

Fig.~\ref{fig7:neon} shows the wavelength-calibration image obtained with a Neon lamp. Wavelength decreases toward the right, while the 38 photonic-lantern outputs are arranged along the vertical axis. The LSF morphology varies both with wavelength and from port to port across the detector.

These considerations motivate a full forward-modeling approach to spectral extraction: the spectral traces on the detector are modeled using empirical LSFs and wavelength solutions derived from calibration data, rather than being extracted with an assumed 1D profile. In wavelength regions where Neon lamp emission lines are absent, the LSF models are obtained by two-dimensional interpolation across the detector. In this way, aberrations across the detector are incorporated directly into the model.


\subsection{Detector nonlinearity and its calibration}\label{sec:detector}



\begin{figure} 
\centering
\includegraphics[width=1\textwidth]{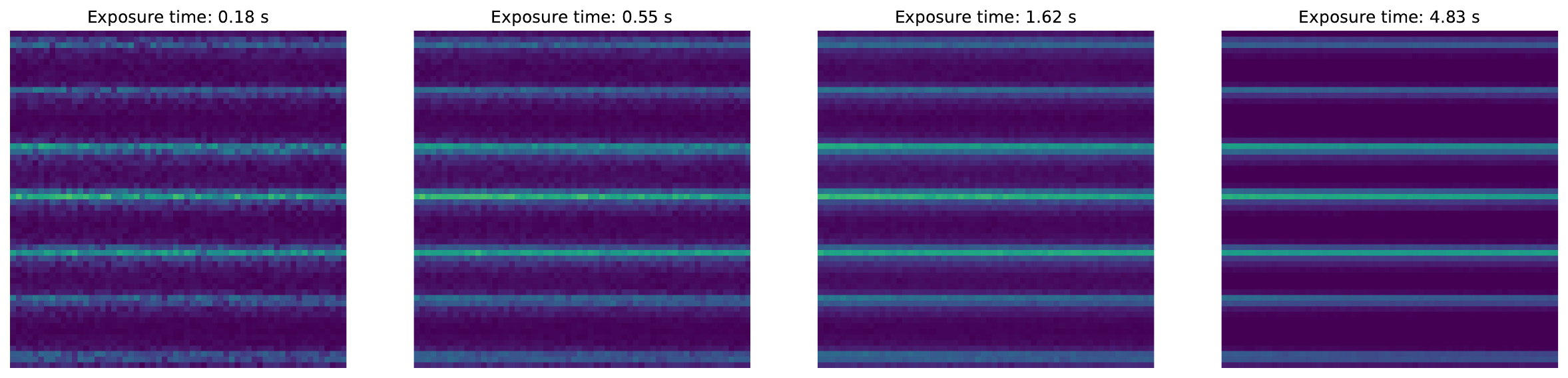}
\caption{A portion of dark-subtracted spectral flat-field images of the Subaru/FIRST-PL detector 
(Orca-Quest qCMOS), normalized by the detector integration time (DIT), for four 
increasing DITs from left to right (0.18, 0.55, 1.62, and 4.83\,s). The total 
integration time is fixed to $\sim$2\,min across all panels, so the total observed 
flux is the same in each image. At short DITs (low flux per frame), the detector 
response is highly nonuniform across pixels; as the DIT increases and pixels enter 
their linear regime, the flat becomes progressively more uniform.}\label{fig7:flat_images}
\end{figure}

\begin{figure}
\centering
    \includegraphics[width=0.75\textwidth]{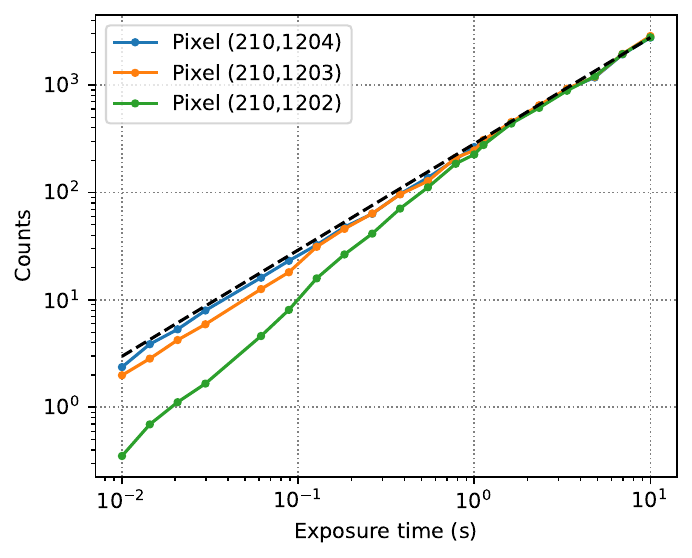}
    \includegraphics[width=0.9\textwidth]{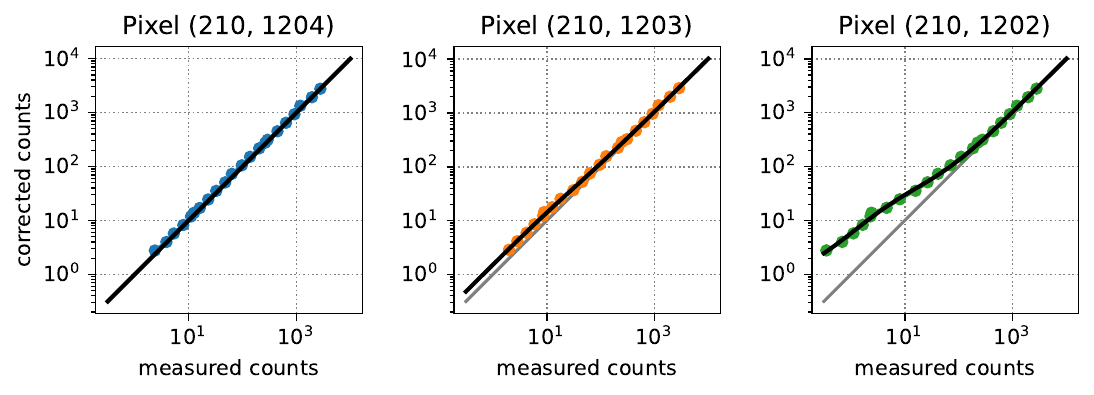}
\caption{\textbf{Top:} Dark-subtracted counts as a function of DIT for three 
adjacent pixels (210, 1202), (210, 1203), and (210, 1204). Dashed lines show 
linear fits to the high-flux regime ($\gtrsim 10^3$\,counts). All three pixels 
exhibit suppressed response at low counts relative to the linear extrapolation, 
with the onset and severity of nonlinearity varying significantly between pixels. 
\textbf{Bottom:} Per-pixel nonlinearity correction curves for the same three pixels, 
showing corrected counts versus measured counts. The gray diagonal indicates perfect 
linearity ($y = x$); the black curves show the empirical correction model fit to 
each pixel.}\label{fig7:nonlinearity_curve}
\end{figure}

The response-map method sidesteps the detector nonuniformity effects to the first order. However, we found a more subtle effect that persists even with the response-map method. 
The FIRST-PL uses an Orca-Quest qCMOS detector from Hamamatsu, a low-read-noise, 
photon-number-resolving sensor increasingly used for high-speed visible imaging 
\cite{strakhov23, lucas24_vampires, layden26}. We found that the detector exhibits 
significant nonlinearity in the low-count regime, where the response is suppressed 
relative to a linear extrapolation from the high-flux regime. Crucially, the degree 
of nonlinearity varies between pixels, which produces apparent nonuniformity in flat-field images 
at low flux levels, while flats obtained at higher flux levels appear significantly more uniform.
Fig.~\ref{fig7:flat_images} demonstrates this: 
dark-subtracted halogen lamp flats normalized by DIT are shown for four increasing 
DITs, with total integration time fixed at $\sim$2 minutes so that photon noise
is negligible compared to the observed structure. At short DITs the flats are highly nonuniform 
--- a structure that reflects true pixel response variations rather than photon 
noise, since the integration time was chosen to make the latter negligible. As the 
DIT increases and pixels enter their linear regime, the flats become progressively 
more uniform. The photoresponse curves of individual pixels confirm this picture: 
the top panel of Fig.~\ref{fig7:nonlinearity_curve} shows dark-subtracted counts 
as a function of DIT for three representative pixels, exhibiting clear 
suppression at low counts, with the onset and severity varying between pixels.

During the $\beta$ CMi observations \cite{kim25}, short integration times were required to freeze 
atmospheric turbulence, placing the observations in this low-flux regime. 
This flux-dependent nonlinearity introduces biases in two ways. First, it can distort spectral shapes ---
because continuum and emission or absorption features correspond to different flux levels,
the apparent widths of spectral features can become biased. 
Second, it can affect spectral-differential
size measurements \cite{porter04, baines06} that the response-map method cannot fully mitigate.
The sharpness of structures in the response maps 
depends on the flux level: in the low-flux regime, stronger nonlinearity 
suppresses low-count pixels relative to their true response, artificially 
sharpening the response map structures.
Because response map sharpness directly determines the inferred spatial 
extent of the source, this bias propagates into size estimates of the 
input scene.
We observed this effect directly in the $\beta$~CMi data: prior to 
correction, size signals across the H$\alpha$ line were noticeably elevated, 
consistent with this nonlinearity bias.

To correct for this, we characterize the photoresponse curve of each pixel 
individually using flat-lamp data obtained across a wide range of DITs. Assuming 
linearity above $\gtrsim$1000 counts, we fit a linear relation to the high-flux 
regime and compute per-pixel correction factors as the ratio of expected to observed 
counts. Because the nonlinear behaviors vary between pixels, two empirical models are 
adopted, both constrained to converge to $y = x$ at high flux:
\begin{equation}
\text{model 1:}\quad y = x \left( 
\frac{1 + a_1/x + a_2/x^2}{1 + a_3/x + a_4/x^2}\right)
\end{equation}
\begin{equation}
\text{model 2:}\quad y = x \left( 
1 + \frac{a_1}{1 + (x/a_2)^{a_3}}\right)
\end{equation}
We found that most pixels could be well fit by one of these two models (bottom 
panels of Fig.~\ref{fig7:nonlinearity_curve}). Once these models are determined from
flat-lamp data, the correction can be applied to observed data. In practice, because 
individual short-exposure 
frames are dominated by photon noise at low flux, the correction is applied to 
averaged frames within each PSF position bin rather than individual frames. After 
applying this correction to the $\beta$~CMi data, the elevated size signals across 
H$\alpha$ were significantly reduced.
This analysis highlights that careful detector characterization is necessary for
accurate spectroastrometric measurements with photonic lantern-fed spectrometers,
and that detector uniformity and linearity may be critical considerations in the
choice of detector for future instruments.

\section{Conclusion}

Photonic lantern spectroastrometry introduces a distinct instrumental artifact landscape compared with traditional slit-based spectroastrometry. The response-map method mitigates some systematics by avoiding direct comparisons of flux across detector pixels, but robust measurements still require careful instrument-specific calibration. For Subaru/FIRST-PL, the dominant practical issues were output-dependent spectral extraction errors and low-count detector nonlinearity. Their mitigation was essential to the on-sky $\beta$ CMi result and is implemented in \texttt{PLred}, providing a practical framework for future photonic lantern-fed spectroastrometric instruments.


\acknowledgments 
This work is supported by the National Science Foundation under Grant Nos. 2109231, 2109232, 2308360, and 2308361. 
The development of SCExAO is supported by the Japan Society for the Promotion of Science (Grant-in-Aid for Research  No. 23340051, 26220704, 23103002, 19H00703, 19H00695 and 21H04998), the Subaru Telescope, the National Astronomical Observatory of Japan, the Astrobiology Center of the National Institutes of Natural Sciences, Japan, the Mt Cuba Foundation and the Heising-Simons Foundation. The development of the CACAO software is supported by the National Science Foundation under award 2410616. E.H. acknowledges funding support for the FIRST project by the French National Research Agency (ANR-21-CE31-0005), E.H. and S.L. acknowledge funding from the project ``Photonics" financed by the ANR program PEPR Origins (ANR-22-EXOR-0005). 

\bibliography{phdbib} 
\bibliographystyle{spiebib} 

\end{document}